\documentclass[12pt, a4paper]{article}

\usepackage[left=2.5cm, right=2.5cm, top=2.0cm, bottom=2.0cm]{geometry}
\usepackage{amsmath, amssymb, graphicx}
\usepackage[utf8]{inputenc}
\usepackage[T2A]{fontenc}
\usepackage[english]{babel}
\usepackage{amsthm}
\usepackage{color}
\usepackage{hyperref}
\hypersetup{colorlinks,citecolor=blue,linkcolor=blue}

\newcommand {\tr} {\text{Tr}}
\newcommand {\bra} [1] {\langle#1|}
\newcommand {\ket} [1] {|#1\rangle}
\newcommand {\avg} [1] {\left\langle#1\right\rangle}

\newcommand {\scal} [2] {\langle#1|#2\rangle}
\newcommand {\ds} {\displaystyle}
\newcommand {\intl} {\int\limits}

\theoremstyle{plain}
\newtheorem{theorem}{Theorem}

\newenvironment{Proof}
{\par\bigskip\noindent{\bf Proof.}} 
{\hfill$\blacksquare$}

\title{A Note on the Eigenstate Thermalization Hypothesis}
\author{Oleg Inozemcev, Igor Volovich} 
\date{}

\begin{document}

\maketitle

Steklov Mathematical Institute of Russian Academy of Sciences

119991, Russia, Moscow, Gubkina str., 8

\vspace{10mm} E-Mail: inozemcev@mi-ras.ru, volovich@mi-ras.ru

\vspace{5cm}
\begin{center}
	\textbf{Abstract}
\end{center}

\begin{center}
	\parbox{12cm}{Eigenstate thermalization hypothesis (ETH) is discussed. We point out that the common formulation of the ETH suffers from the mixing of random and deterministic variables. We suggest a modified formulation of the ETH which includes only deterministic variables. We also show that this formulation of the ETH implies thermalization as well.}
\end{center}

\newpage
\section{Introduction}
Recent advancement in the field of thermalization in quantum systems has provided a better understanding of these questions (see comprehensive overviews \cite{review1,review2,Deutsch_ETH}). Now it is known as \textit{eigenstate thermalization hypothesis (ETH)}. This hypothesis states that all the energy eigenstates in an energy shell represent thermal equilibrium.

These ideas go back to von Neumann's paper \cite{Neumann29} of 1929 where the main notions of statistical mechanics were re-interpreted in a quantum-mechanical way, the ergodic theorem and the H-theorem were formulated and proved. He also noted that, when discussing thermalization in isolated quantum systems, one should focus on physical observables as opposed to wave functions or density matrices describing the entire system.

The next important step was made in 1991 by Deutsch \cite{Deutsch1}. In his paper based on the random matrix theory (RMT), the relationship between diagonal matrix elements of an observable and microcanonical averages was revealed.

In later important works, Srednicki provided a generalization of the RMT prediction. In 1994 for a gas of hard spheres at high energy and low density, he showed \cite{Sredn1} that each energy eigenstate which satisfies Berry’s conjecture predicts a thermal distribution for the momentum of a single constituent particle. He referred to this remarkable phenomenon as \textit{eigenstate thermalizatio}n. The current formulation of ETH was given by Srednicki in 1999 \cite{Sredn3}.

We point out that the common formulation of the ETH suffers from the mixing of random and deterministic variables. We suggest a modified formulation of the ETH which has only deterministic variables. We also show that this formulation of the ETH implies thermalization as well.

In the next section the deterministic formulation of the ETH is presented. Following \cite{Sredn3}, in section 3 we provide a detailed proof that deterministic formulation of the ETH entails thermalization as well.

\vspace{1cm}
\section{Common formulation of the ETH}
Consider a bounded, isolated, many-body quantum system with $N$ degrees of freedom, where $N \gg 1$. Since the system is bounded, the energy eigenvalues are discrete, and since it is isolated, its time evolution is governed by the Schrodinger equation \cite{Sredn3}. 

Denote energy eigenvalues as $E_\alpha$ and corresponding eigenstates as $\ket{\alpha}$, so that
\begin{equation}\label{}
	\hat{H}\ket{\alpha} = E_\alpha\ket{\alpha}.
\end{equation}
The state of the system at any time $t$ is then given by 
\begin{equation}\label{state_t}
	\ket{\psi_t} = \sum_\alpha c_\alpha \,e^{-iE_\alpha t}\ket{\alpha},
\end{equation}
where as usual
\begin{equation}\label{}
	\scal{\psi_t}{\psi_t} = \sum_\alpha |c_\alpha|^2 = 1.
\end{equation}

Let an observable of interest be a smooth function of the classical coordinates and momenta which has no explicit dependence on $\hbar$. The hermitian operator $\hat{A}$ corresponds to this observable, and
\begin{equation}\label{}
	A_{\alpha\beta} := \bra{\alpha}\hat{A}\ket{\beta}
\end{equation}
are its matrix elements.

According to Srednicki \cite{Sredn2}, quantum chaos theory (see, for example, \cite{GianVorZinn}-\cite{OhyaVol} and refs therein) predicts that the matrix elements $A_{\alpha\beta}$ 
are given by
\begin{equation}\label{anzatz0}
	A_{\alpha\beta} = \mathcal{A}(E_\alpha)\delta_{\alpha\beta} + \hbar^{(N-1)/2}R_{\alpha\beta},
\end{equation}
where \\
$\mathcal{A}(E)$ is a smooth function of energy whose leading term in the $\hbar$ expansion is $O(1)$,\\
$\delta_{\alpha\beta}$ is the Kronecker delta, \\
$R_{\alpha\beta}$ are also $O(1)$ at leading order, and their values are characterized by a smooth distribution, often assumed to be gaussian. 

In a later paper \cite{Sredn3} Srednicki suggested the following formula for the matrix elements of A in the energy eigenstate basis
\begin{equation}\label{anzatzSr}
	A_{\alpha\beta} = \mathcal{A}(E)\delta_{\alpha\beta} + e^{-S(E)/2}f(E,\omega)R_{\alpha\beta}
\end{equation}
where \\
$E := \frac{1}{2}(E_\alpha+E_\beta)$, \\
$\omega := E_\alpha-E_\beta$, \\
$\mathcal{A}(E)$ is a smooth function which can be related to the expectation value of the canonical ensemble at energy $E$ (see Theorem 1 below), \\ 
$S(E)$ is the thermodynamic entropy at energy $E$, given by
\begin{equation}\label{entropy}
	e^{S(E)} = E \sum_\alpha \delta_\varepsilon(E-E_\alpha), 
\end{equation}
where $\delta_\varepsilon(x)$ is a Dirac delta function that has been smeared just enough to render $S(E)$
monotonic, \\
$f(E,\omega)>0$ is an even function of $\omega$ and a smooth function of both arguments, \\
$R_{\alpha\beta}\in \mathbb{C}$ is a random variable which varies erratically with $\alpha$, $\beta$, \\
$\mathbb{E}\left[\text{Re}(R_{\alpha\beta})\right] = \mathbb{E}\left[\text{Im}(R_{\alpha\beta})\right] = 0$, $\text{Var}\left[\text{Re}(R_{\alpha\beta})\right] = \text{Var}\left[\text{Im}(R_{\alpha\beta})\right] = 1$,\\
$N \gg 1$ and the entropy is supposed to be extensive, i.e. 
\begin{equation}\label{extens}
	S(E,N) = Ns(E/N) + O(\log N).
\end{equation}

Formula \eqref{anzatzSr} is referred to as the \textit{ETH-ansatz}, or \textit{Srednicki's ansatz}.

\bigskip
There is a point about ETH written in the form \eqref{anzatzSr} (and also \eqref{anzatz0}). For a given (non-random) Hamiltonian, the RHS of \eqref{anzatzSr} is random while the LHS is not. In this case equality \eqref{anzatzSr} is not quite correct. 

The ordinary arguments are as follows. In \cite{Deutsch1,Deutsch2,Reimann_Dappr} Hamiltonians are taken in the form
\begin{equation}\label{Ham}
	\hat{H} = \hat{H}_0 + \hat{V},
\end{equation}
where $\hat{H}_0$ is an “unperturbed” part, $\hat{V}$ is a weak “perturbation”. One can think of $\hat{H}_0$ as describing an ideal gas in a box and $\hat{V}$ as describing two-particle interactions. Instead of adding in these interactions explicitly, $\hat{V}$ is modeled by a random matrix from a certain random matrix ensemble with statistical properties which imitate well the main features of the perturbation $\hat{V}$ (band structure, sparsity etc). That is the many-body Hamiltonian $\hat{H}$ resembles a random matrix. Hence, its eigenstates and matrix elements are random as well.

However, for a deterministic Hamiltonian, the formula \eqref{anzatzSr} is not true since there is the deterministic value in the LHS, but the RHS includes the random variable $R_{\alpha\beta}$.

\vspace{1cm}
\section{Deterministic formulation of the ETH}

For a given non-random Hamiltonian, we suggest to write the ETH in the following form
\begin{equation}\label{anzatz2}
	A_{\alpha\beta} = \mathcal{A}(E)\delta_{\alpha\beta} + e^{-S(E)/2}g_{\alpha\beta}(E,\omega),
\end{equation}
where $g_{\alpha\beta}(E,\omega)$ is a bounded complex-valued (non-random) function smooth of both arguments, and $N \gg 1$.

Now we make sure that the two important assertions (see theorems 1, 2 below) related to the ETH in the form of \eqref{anzatzSr} still holds for the ETH in the form of \eqref{anzatz2}.

Recall that the canonical thermal average is
\begin{equation}\label{canon_avg}
	\avg{A}_T = \frac{\tr\,e^{-H/T}\!A} {\tr\,e^{-H/T} },
\end{equation}
where $\ds T=\frac{dE}{dS}$ is the temperature.

It was argued in \cite{Sredn3} that $\mathcal{A}(E)$ approximately equals to $\avg{A}_T$, namely the following theorem holds.

\begin{theorem}
	Suppose that deterministic formulation \eqref{anzatz2} of the ETH holds. Then for large degrees of freedom $N$
	\begin{equation}\label{calAE}
		\mathcal{A}(E) = \avg{A}_T + O(N^{-1}),
	\end{equation}
	where $\mathcal{A}(E)$ and $\avg{A}_T$ are defined above.
\end{theorem}

We reproduce here a detailed proof of this theorem with the ETH-ansatz in the form of \eqref{anzatz2} instead of \eqref{anzatzSr}.

\begin{Proof}
The numerator in \eqref{canon_avg} is
\begin{align}
	Z\avg{A}_T &:= \tr \left[ \exp\left(-\dfrac{H}{T}\right)\!A \right] = 
	\sum_\alpha \left[ \exp\left(-\dfrac{E_\alpha}{T}\right) A_{\alpha\alpha} \right] = \\
	&= \sum_\alpha \left[ \int \exp\left(-\dfrac{u}{T}\right) \delta(u-E_\alpha) 
	\,du \,A_{\alpha\alpha} \right] = \\
	&= \int \exp\left(-\dfrac{u}{T}\right) \sum_\alpha \delta(u-E_\alpha) A_{\alpha\alpha} \,du,
\end{align}
where $u\in(0,+\infty)$ is the integration variable.
Substitute ETH \eqref{anzatz2} into the last expression

\begin{align}
	Z\avg{A}_T &= \int \exp\left(-\dfrac{u}{T}\right) \sum_\alpha \delta(u-E_\alpha) 
		\left[\mathcal{A}(E_\alpha) + O\left(e^{-S(E_\alpha)/2}\right)\right] du = \\
	&= \int \exp\left(-\dfrac{u}{T}\right) \left[\mathcal{A}(u) + O\left(e^{-S(u)/2}\right)\right] \sum_\alpha \delta(u-E_\alpha) \,du. \label{l1}	
\end{align}

From \eqref{entropy} we have
\begin{equation}\label{delta}
\sum_\alpha \delta(E-E_\alpha) \approx \frac{e^{S(E)}}{E}. 
\end{equation}
Substitute \eqref{delta} into \eqref{l1}
\begin{align}
	&Z\avg{A}_T = \\
	&= \int \frac{1}{u} \exp\left(S(u)-\dfrac{u}{T}\right) \mathcal{A}(u) \,du + 
		\int \frac{1}{u} \exp\left(S(u)-\dfrac{u}{T}\right) O\left(e^{-S(u)/2}\right) du.
\end{align}

Similarly, the denominator in \eqref{canon_avg} is 
\begin{equation}\label{}
	Z := \tr\,e^{-H/T} = \int \frac{1}{u} \exp\left(S(u)-\frac{u}{T}\right)du.
\end{equation}
Then \eqref{canon_avg} takes the form
\begin{equation}\label{ints}
	\avg{A}_T = \frac{\ds\int_0^\infty \frac{1}{u} \: e^{S(u)-u/T} \mathcal{A}(u)\,du}
	{\ds\int_0^\infty \frac{1}{u} \: e^{S(u)-u/T}du} + 
	\frac{\ds\int_0^\infty \frac{1}{u} \: e^{S(u)-u/T} O\!\left(e^{-S(u)/2}\right) du} 
	{\ds\int_0^\infty \frac{1}{u} \: e^{S(u)-u/T}du}.	
\end{equation}

Since the entropy is extensive for large N (see \eqref{extens}), the integrals in \eqref{ints} can be evaluated by Laplace's method (see, for example, \cite{Fedoryuk,Wong}). We find that

\begin{equation}\label{A_t}
	\avg{A}_T = \mathcal{A}(u) + O\!\left(e^{-S/2}\right) + O(N^{-1}) = \mathcal{A}(u) + O(N^{-1}),
\end{equation}
where $u$ satisfies $\ds \frac{\partial S}{\partial u} = \frac{1}{T}$. From \eqref{A_t} we get finally
\begin{equation}\label{}
	\mathcal{A}(u) = \avg{A}_T + O(N^{-1}),
\end{equation}
which is the statement of the theorem.
\end{Proof}
\bigskip

Using ETH \eqref{anzatzSr}, thermalization of eigenstates was proposed in \cite{Sredn3}. Note that, in fact, randomness of $R_{\alpha\beta}$ in \eqref{anzatzSr} is not needed for the proof of assertion about thermalization.

\begin{theorem}
	Suppose that deterministic formulation \eqref{anzatz2} of the ETH holds.
	Let the state of a bounded and isolated quantum system is given by
	\[ \ket{\psi_t} = \sum_\alpha c_\alpha e^{-iE_\alpha t} \ket{\alpha}, \]
	where $E_\alpha$ are non-degenerate energy eigenvalues corresponding to energy eigenstates $\ket{\alpha}$.
	Let further define \\
	$\ds A_t := \bra{\psi_t}\hat{A}\ket{\psi_t} = 
	\sum_{\alpha\beta} c_\alpha^* c_\beta\,e^{i(E_\alpha-E_\beta)t} A_{\alpha\beta}$ is the expectation value of an observable $A$, \\
	$\ds \overline{A} := \lim_{\tau\to\infty} \frac{1}{\tau} \int_0^\tau A_t\,dt$ is its infinite time average (Сеsarо mean), \\
	$\ds E := \sum_\alpha |c_\alpha|^2E_\alpha$ is the expectation value of the total energy, \\
	$\ds \varDelta^2 := \sum_\alpha |c_\alpha|^2 (E_\alpha-E)^2$ is a quantum uncertainty of the total energy expectation value, and assume that
	\begin{equation}\label{quant_uncert}
		\varDelta^2 \left|\frac{\mathcal{A}''(E)}{\mathcal{A}(E)}\right| \ll 1.
	\end{equation}
	Then for large $N$ \\
	1) the infinite time average of $A_t$ is equal to its equilibrium value \eqref{canon_avg} at the appropriate temperature
	\[ \overline{A} = \avg{A}_T + O\!\left(\varDelta^2\right) + O\!\left(N^{-1}\right), \]
	2) the fluctuations of $A_t$ about $\overline{A}$ are very small
	\[ \overline{(A_t-\overline{A})^2} = O\!\left(e^{-S}\right). \]
\end{theorem}

One can see that the arguments in \cite{Sredn3} remains valid if ETH \eqref{anzatzSr} is replaced with \eqref{anzatz2}. Here we reproduce this proof in details.

\begin{Proof}
Expectation value $A_t$ can be written as
\begin{align}
	\label{At1} A_t \equiv \bra{\psi(t)}\hat{A}\ket{\psi(t)} = 
	\sum_{\alpha,\beta} c_\alpha^* c_\beta\,e^{i(E_\alpha-E_\beta)t} A_{\alpha\beta} = \\ 
	\label{At2} = \sum_\alpha |c_\alpha|^2 A_{\alpha\alpha} + 
	\sum_{\alpha,\beta\neq\alpha} c_\alpha^* c_\beta\,e^{i(E_\alpha-E_\beta)t} A_{\alpha\beta}. 
\end{align}

Since $E_\alpha$ are non-degenerate energy eigenvalues then, substituting this $A_t$ \eqref{At2} into the definition of $\overline{A}$, we get
\begin{align}
	\overline{A} &= \lim_{\tau\to\infty} \frac{1}{\tau} \intl_0^\tau
	\left( \sum_\alpha |c_\alpha|^2 A_{\alpha\alpha} + 
	\sum_{\alpha,\beta\neq\alpha} c_\alpha^* c_\beta\,e^{i(E_\alpha-E_\beta)t} A_{\alpha\beta}\right)dt =\\ \label{Ces_mean1}&= \sum_\alpha |c_\alpha|^2 A_{\alpha\alpha}.
\end{align}

Now we use the ETH-ansatz, namely substitute \eqref{anzatz2} into \eqref{Ces_mean1}
\begin{align}
	\overline{A} &= \sum_\alpha |c_\alpha|^2 \left[ \mathcal{A}(E_\alpha) + 
		O\left(e^{-S(E_\alpha)/2}\right) \right] = \\
	&= \sum_\alpha |c_\alpha|^2 \mathcal{A}(E_\alpha) + 
		\sum_\alpha |c_\alpha|^2 O\left(K_\alpha e^{-S(E_\alpha)/2}\right) = \\
	&= \sum_\alpha |c_\alpha|^2 \mathcal{A}(E_\alpha) + 
		\sum_\alpha |c_\alpha|^2 O\left(e^{-S(E)/2}\right) = \\
	&= \sum_\alpha |c_\alpha|^2 \mathcal{A}(E_\alpha) + O\left(e^{-S(E)/2}\right). \label{Ces_mean2}
\end{align}

Note that it makes no difference whether we use ETH-ansatz in the form of \eqref{anzatzSr} or in the form of \eqref{anzatz2}.

Expansion of the smooth function $\mathcal{A}(E_\alpha)$ into a Taylor series around $E$ is
\begin{equation}\label{}
	\mathcal{A}(E_\alpha) = \mathcal{A}(E) + \mathcal{A'}(E)\,(E_\alpha-E) + \frac{1}{2}\mathcal{A''}(E)\,(E_\alpha-E)^2 + O[(E_\alpha-E)^3].
\end{equation}
Substitute this expansion into \eqref{Ces_mean2}:
\begin{align}
	\overline{A} &= \sum_\alpha |c_\alpha|^2 \mathcal{A}(E) + 
	\sum_\alpha |c_\alpha|^2 \mathcal{A}'(E) (E_\alpha-E) + \\
	&+ \sum_\alpha |c_\alpha|^2 \frac12\mathcal{A''}(E)\,(E_\alpha-E)^2 + 
	\sum_\alpha O[(E_\alpha-E)^3] + O(e^{-S(E)/2}) = \\
	&= \mathcal{A}(E) + \mathcal{A}'(E)\sum_\alpha |c_\alpha|^2E_\alpha - 
	E\mathcal{A}'(E)\sum_\alpha |c_\alpha|^2 + \\
	&+ \frac12\mathcal{A''}(E)\sum_\alpha |c_\alpha|^2(E_\alpha-E)^2 + 
	O[(E_\alpha-E)^3] + O(e^{-S(E)/2}) = \\
	&= \mathcal{A}(E) + O(\varDelta^2) + O(e^{-S(E)/2}). \label{Ces_mean3}
\end{align}
Now we use theorem 1, namely substitute \eqref{calAE} into \eqref{Ces_mean3}:
\begin{equation}\label{}
	\overline{A} = \avg{A}_T + O\!\left(N^{-1}\right) + O\!\left(\varDelta^2\right) + O(e^{-S(E)/2}) =
	\avg{A}_T + O\!\left(N^{-1}\right) + O\!\left(\varDelta^2\right),
\end{equation}
which is the first statement of the theorem.

By definition
\begin{equation}\label{fluct}
	\overline{(A_t-\overline{A})^2} = \lim_{\tau\to\infty} \frac{1}{\tau} \intl_0^\tau 
	(A_t - \overline{A})^2dt = 
	\lim_{\tau\to\infty} \frac{1}{\tau} \intl_0^\tau A_t^2\,dt - \overline{A}^2.
\end{equation}
From \eqref{At1} we get 
\begin{align}
	A_t^2 = \sum_{\alpha,\beta,\gamma,\delta} c_\alpha^* c_\beta c_\gamma^* c_\delta \exp[i(E_\alpha-E_\beta+E_\gamma-E_\delta)t] A_{\alpha\beta}A_{\gamma\delta} = \\
	\label{At_squar} = \sum_{\alpha,\gamma} |c_\alpha|^2 |c_\gamma|^2 A_{\alpha\alpha}A_{\gamma\gamma} + 	
	\sum_{\alpha,\beta\neq\alpha} |c_\alpha|^2 |c_\beta|^2 A_{\alpha\beta}A_{\beta\alpha} + (\text{other terms}).
\end{align}
From \eqref{Ces_mean1} we obtain 
\begin{equation}\label{Ces_mean_squar}
	\overline{A}^2 = \sum_{\alpha,\gamma} |c_\alpha|^2 |c_\gamma|^2 A_{\alpha\alpha}A_{\gamma\gamma}
\end{equation}
Substitute \eqref{At_squar} and \eqref{Ces_mean_squar} into \eqref{fluct}: 
\begin{align*}
	\overline{(A_t-\overline{A})^2} &= 
	\sum_{\alpha,\gamma} |c_\alpha|^2 |c_\gamma|^2 A_{\alpha\alpha}A_{\gamma\gamma} + 	
	\sum_{\alpha,\beta\neq\alpha} |c_\alpha|^2 |c_\beta|^2 A_{\alpha\beta}A_{\beta\alpha} - \\
	&-\sum_{\alpha,\gamma} |c_\alpha|^2 |c_\gamma|^2 A_{\alpha\alpha}A_{\gamma\gamma} =
	\sum_{\alpha,\beta\neq\alpha} |c_\alpha|^2 |c_\beta|^2 |A_{\alpha\beta}|^2 \leqslant \\
	&\leqslant \max_{\alpha\neq\beta}|A_{\alpha\beta}|^2 \sum_{\alpha,\beta} |c_\alpha|^2 |c_\beta|^2 =
	\max_{\alpha\neq\beta}|A_{\alpha\beta}|^2. 	
\end{align*}
Using the ETH \eqref{anzatz2}, we get finally
\begin{equation}
	\overline{(A_t-\overline{A})^2} \leqslant \max_{\alpha\neq\beta}|A_{\alpha\beta}|^2 = 
	\max_{\alpha\neq\beta} \left\lbrace O\!\left(e^{-S([E_\alpha+E_\beta]/2)}\right) \right\rbrace =
	O\!\left(e^{-S(E)}\right),
\end{equation}
which is the second statement of the theorem.
\end{Proof}
\bigskip

\vspace{1cm}
\section{Conclusion}
The common formulation of the eigenstate thermalization hypothesis suffers from the mixing of random and deterministic variables. That is for a given (non-random) Hamiltonian, the RHS of \eqref{anzatzSr} is random while the LHS is not.

We have suggested deterministic formulation \eqref{anzatz2} of the ETH with non-random variables only. 
Following \cite{Sredn3}, we have considered detailed proofs of two theorems related to the ETH, namely theorem about canonical thermal average (Theorem 1) and theorem about thermalization of a quantum system (Theorem 2). 
It is easy to see that to prove these theorems one does not need the ETH in the form of \eqref{anzatzSr}, it is sufficient to use the deterministic formulation \eqref{anzatz2} of the ETH.

\end{document}